# Hyperbolic phonon-polaritons in boron nitride for near-field optical imaging and focusing

Peining Li[1], Martin Lewin[1,2], Andrey V. Kretinin,[3] Joshua D. Caldwell[4]*, Kostya S. Novoselov,[3] Takashi Taniguchi,[5] Kenji Watanabe,[5] Fabian Gaussmann[2], Thomas Taubner[1,2]*

1 Institute of Physics (IA), RWTH Aachen University, Aachen 52056, Germany

2 Fraunhofer Institute for Laser Technology ILT, 52074 Aachen, Germany

3 School of Physics and Astronomy, University of Manchester, Oxford Rd, Manchester, UK

4 U.S. Naval Research Laboratory, 4555 Overlook Ave, S.W., Washington, D.C. USA

5 National Institute for Materials Science, 1-1 Namiki, Tsukuba, Ibaraki 305-0044, Japan

*Correspondence to: joshua.caldwell@nrl.navy.mil and taubner@physik.rwth-aachen.de

**Natural hyperbolic materials (NHMs) exhibit sub-diffractional, highly directional, volume-confined polariton modes. Here we report that hyperbolic phonon polaritons (HPs) allow for a flat slab of hexagonal boron nitride (hBN) to enable novel near-field optical applications, including unusual imaging phenomenon (such as an enlarged reconstruction of investigated objects) and sub-diffractional focusing. Both the enlarged imaging and the super-resolution focusing are explained based on the volume-confined, wavelength dependent propagation angle of HPs. With advanced infrared nanoimaging techniques and state-of-art mid-infrared laser sources, we have succeeded in demonstrating and visualizing**

**these unexpected phenomena for the first time in both Type I and Type II hyperbolic conditions, with both occurring naturally within hBN. These efforts have provided a full and intuitive physical picture for the understanding of the role of HPs in near-field optical imaging, guiding, and focusing applications.**

**Introduction**

The propagation of sub-diffractional waves in hyperbolic media[1] enables many unusual optical possibilities such as hyperlensing[2,3,4], negative refraction[5,6], enhanced quantum radiation[7], nanolithography[8] and sub-diffractional resonators[9,10]. Very recently, it was demonstrated that the highly directional propagation of volume-confined, hyperbolic polaritons (HPs) is key for these sub-diffractional phenomena[8,10,11]. Their directionality derives from the sign and magnitude of the two principal (in- and out-of-plane) components of the dielectric-permittivity tensor ($\vec{\varepsilon} = diag[\varepsilon_{xx}, \varepsilon_{yy}, \varepsilon_{zz}]$), which have opposite signs in hyperbolic materials. The propagation angle $\theta$ (e.g. the angle between the Poynting vector and the z-axis) of the HPs in hyperbolic media can be roughly approximated as[10],

$$\theta = \pi/2 - \arctan\left(\sqrt{\varepsilon_z(\omega)}/i\sqrt{\varepsilon_t(\omega)}\right) \quad (1)$$

where $\varepsilon_t = \varepsilon_{xx} = \varepsilon_{yy}$ and $\varepsilon_z = \varepsilon_{zz}$ are the in- and out-of-plane dielectric permittivities of the hyperbolic medium, respectively. Therefore, by controlling the ratio of the two principal dielectric components, the propagation direction of the HPs can be tuned.

Until very recently, hyperbolic media have been explored through man-made hyperbolic metamaterial structures, such as metal-dielectric multilayers[4,5], nanowire [6,12] or nano-pyramid arrays[9] embedded within a dielectric medium. In hyperbolic metamaterials (HMMs), the effective dielectric permittivities are determined by the geometric parameters of their subwavelength unit cells[1]. As such, the maximum wavevector **k** that can be induced to propagate through the material is limited by the size of the artificial unit cell. This in turn limits the degree of optical confinement and spatial resolution that can be realized. Furthermore, the high losses associated with noble metals[13,14,15] used in man-made hyperbolic metamaterial structures result in short propagation lengths, quite broad resonance linewidths and in terms of hyperlens designs, low transmission efficiency.

During the search for better "plasmonic" materials[14,15], polar dielectrics capable of supporting phonon-polaritons such as silicon carbide[16,17,18,19,20] and hBN[10,11,21] have been demonstrated as superior alternatives to metals at mid-infrared to THz frequencies. Interestingly, many phonon-resonant materials such as quartz[22], zinc oxide[23], calcite[24] and hBN[10,11,21] are also natural hyperbolic materials[25,26,27] (NHMs). These NHMs support HPs within homogeneous crystals with atomic-scale unit cells, thus the upper limit on the highest propagating wavevectors **k** associated with artificial metal-dielectric HMMs is no longer an issue. Instead, photonic confinement within tiny volumes in the few nanometer range becomes possible. This was recently demonstrated by Dai *et al.* [11] where surface phonon polariton propagation within a 3 monolayer (<1 nm) thin flake of hBN was reported. It is the propagation of such high-**k** fields that are scattered off or launched from deeply sub-diffractional objects that is at the heart of super-resolution imaging. These benefits are also coupled with a drastic reduction in the optical losses compared to HMMs, which results in improved performance, i.e. higher field confinement[10,11] and improved image resolution.

In contrast to HMMs reported to date, hBN offers the additional functionality of sub-wavelength imaging in different spectral regions through the presence of two separate spectral bands that exhibit inverted hyperbolic response, making this an ideal material for exploring the basic phenomenon of hyperbolic polaritons. These two regimes are referred to as the lower and upper Reststrahlen bands[10,11], where this term refers to the spectral range between the longitudinal (LO) and transverse (TO) optic phonons of a polar crystal where a negative real part of the dielectric function is observed. The presence of two bands results from the highly anisotropic crystal structure of hBN, where both a,b- and c-axis oriented optical phonons are supported and are widely separated in frequency[28]. These two bands not only exhibit hyperbolic behavior, but the crystal axis featuring negative real permittivity is inverted, thus the lower and upper bands offer Type-I (Re($\varepsilon_t$) > 0 and Re($\varepsilon_z$) < 0 in ~760 < $\omega$ < 825 cm$^{-1}$) and Type-II (Re($\varepsilon_t$) < 0 and Re($\varepsilon_z$) > 0 in ~1360 < $\omega$ < 1610 cm$^{-1}$) hyperbolic response, respectively. The inversion of the signs of the dielectric function results in novel behavior, such as a negative (positive) z-component of the group velocity in the upper (lower) Reststrahlen bands. This results in unique phenomenon such as higher order resonance modes occurring at lower (higher) frequency. Until now, due to a lack of a homogeneous material exhibiting both types of hyperbolicity, a comprehensive study of the impact of these two unique regimes for nanoimaging and super-resolution focusing have not been experimentally probed, with Caldwell et al. providing the only prior study comparing the unique behaviors of these two regimes, albeit within the context of three-dimensionally confined cavities. In addition to providing a homogeneous medium exhibiting both Type I and II hyperbolicity, hBN also exhibits much lower losses (higher efficiencies) than plasmonic materials, with the imaginary part of the dielectric function , i.e., Im($\varepsilon_t$) ~ 0.1 for Re($\varepsilon_t$) = −1 at $\omega$ = 809 cm$^{-1}$ and Im($\varepsilon_z$) ~ 0.1 for Re($\varepsilon_z$) = −1 at $\omega$ = 809 cm$^{-1}$, which is crucial for realizing extended propagation and detection of high-**k** polariton modes (deeply sub-diffractional confinement)[10,11,17].

The dielectric permittivities are also highly dispersive in these two regions, giving rise to a frequency-dependent ratio between $\varepsilon_t$ and $\varepsilon_z$. Thus, because of the presence of both Type I and II hyperbolicity, from Eq. (1) we can predict that the propagation angle $\theta$ will be an increasing (decreasing) function of the frequency in the lower (upper) Reststrahlen bands. As shown in Fig. 1 a and b, as the frequency is increased, $\theta$ increases from about 16° to 87° ($\Delta\theta \approx 71°$) within the Type I lower Reststrahlen band, while it decreases from $\theta \approx 83°$ to $\theta \approx 2°$ ($\Delta\theta \approx 81°$) in the Type II upper Reststrahlen band. Here, we experimentally verify this inversion of the propagation angle within the two spectral bands, providing near-field imaging of sub-diffractional objects under a thin slab of hBN. This investigation also enabled the experimental demonstration of the unusual imaging properties of both Type I and II hyperbolic media, thereby providing the physical description necessary to realize and optimize near-field imaging using NHMs.

**Theory and simulations**

This frequency-dependent tuning of the angular HP propagation in hBN and the suitability of this material for near-field imaging can be easily quantified and visualized via two-dimensional (2D) numerical calculations. For this, we consider a 0.3 μm wide gold stripe on a Si substrate with an hBN cover layer (the thickness $h = 1$ μm) illuminated by a $p$-polarized plane wave incident from the top. Simulated electric-field distributions ($|\mathbf{E}_z|$) at six typical operation frequencies are presented in Figs. 1c-h, and demonstrate the directional nature of the HPs. Here the high-**k** fields scattered from the edges of the embedded gold stripe are induced to propagate within the hBN flake (as marked in Fig. 1d), with the angle of propagation being directly dependent upon the frequency of operation. As stated previously[1-4], in the absence of the hyperbolic dispersion, such high-**k** modes would be evanescent (i.e. decay rapidly) within the medium. Each edge of the Au stripe excites two sub-diffractional HPs that propagate at the angles +/- $\theta$. This frequency dependent propagation angle quantitatively agrees with the analytical predictions based on Eq. (1) (see Fig. 1a and b), therefore the propagation angle can be predicted via the simple ratio of the extraordinary (z-axis) and ordinary (x-y plane) components of the anisotropic dielectric function of hBN.

The anticipated super-resolution imaging performance of the hBN slab is directly tied to the propagation angle, as demonstrated in Figs. 1c-h, which as noted above is directly dependent upon the type of hyperbolicity. For instance, at wavelengths with very low propagation angles (Figs. 1 c and f), the image of the sub-diffraction Au stripe is nearly perfectly restored on the top surface of the hBN, similar to near-field superlensing[29,30,31]. Namely, the restored width d' is nearly identical to the actual width $d$ of the investigated stripe (e.g. d' ≈ 309 nm, around $\lambda$/40-resolution at $\omega$ = 778.2 cm$^{-1}$ or λ=12.85 μm). However, as the angle is increased (Figs, 1d,e for the lower and g,h for the upper Reststrahlen bands), an enlarged outline of the object image is obtained with the width $D(\omega)= d + 2 h \tan\theta (\omega)$. Note that due to the inversion of the dependence of $\theta$ on between the two spectral bands, superlensing-type response is observed at low and high frequencies within the lower and upper Reststrahlen bands, respectively, while the enlarged imaging behavior is observed at high and low frequencies.

This novel enlargement can be clearly observed in the more practical three-dimensional (3D) cases. As shown in Fig. 2a, the 0.6-μm-diameter gold disc is once again perfectly restored at very shallow angles; however, at larger propagation angles a double concentric ring-like field distribution is recorded in the near-field (Fig. 2b), rather than a direct replication of the original field distribution of the object. This enlarged pattern results from the frequency-dependent propagation angle of the HPs (cone-like shape for 3D case, see the sketch in Fig. 2c). This unexpected phenomenon is also found for other shapes (like a Au square and stripe) as shown in Figs. d-i. Intriguingly, the trace of the HP cones reconstructs the enlarged and slightly distorted pattern that is still able to identify the outline (shape) information of the object. Quantitatively, through recording the HP-reconstructed outline ($D(\omega)$), we can extract the geometric size and shape from the strict relationship of $D(\omega) = d + 2h\tan\theta(\omega)$. These results suggest a new HP-based imaging scenario, which will be verified and visualized by our experiments below.

**Experimental results in Type I hyperbolic band**

The predicted imaging mechanism is verified by our experiments first in the Type I hyperbolic band, as presented in Fig. 3. A schematic of the experimental setup is provided in Fig. 3a where the images restored by the hBN layer are recorded using a scattering-type scanning near-field optical microscope (s-SNOM). We use a 0.15-μm-thick exfoliated hBN flake to image the underlying, 30 nm tall, gold nanodiscs with 0.3-μm diameter and 1.3-μm center-to-center separation. The metallic tip of the s-SNOM is illuminated by a home-built, tunable broadband infrared laser[32,33] with the peak position of the laser spectrum (inset in Fig. 3a) matched to the lower, Type-I hyperbolic region of hBN (760 cm$^{-1}$ < $\omega$ < 825 cm$^{-1}$). Both the optical and topographic information at the top surface of the hBN layer are collected simultaneously (details in Methods). In the obtained topographic image (Fig. 3b), the gold discs are masked by the covering hBN layer, while in the broadband-SNOM image (Fig. 3c), all three nanodiscs are clearly resolved with bright contrast (i.e. high signal-to-background ratio). To ensure that the imaging is indeed due to the hyperbolic nature of the hBN slab, a control image was also collected at a frequency outside of the lower Reststrahlen band using a $CO_2$ laser at $\omega$ = 952 cm$^{-1}$, where both components of the dielectric function are positive (Re($\varepsilon_t$) = 8.8 and Re($\varepsilon_z$) = 2). As shown in Fig. 3e, in contrast to the hyperbolic case, only weak features of the discs are observed through the thin hBN layer in the control experiment. This is further demonstrated by the line profiles taken across the two discs for both cases presented in Fig. 3f, with a dramatic enhancement in the imaging efficiency observed in the hyperbolic regime. Such an effect would be further amplified within thicker hBN slabs, whereby any structural morphology would be totally lost, but the near-field imaging properties retained. The hyperbolic images also provided a narrower FWHM (full width at half maximum) of about 0.5 μm in addition to the dramatically improved contrast (signal-to-background ratio). Considering the resolved deeply subwavelength optical FWHM (~ $\lambda/24$), this comparison clearly confirms the improved near-field imaging by the hBN layer.

Although the broadband s-SNOM image shown in Fig. 3c is able to resolve the discs, it does not reflect the actual field distribution of the image due to its detection scheme[33]. Because of the broadband light source used, the detection is not monochromatic as in the simulations presented in Fig. 1 and 2. Instead, the collected images are the superposition of all the different frequency components of the broadband laser and therefore the concentric-ring field distributions are not directly observed. Further, depending on the chosen interference phase difference (constructive or destructive phases, involving the position of a reference mirror in our setup, see Methods), it can also show reversed imaging contrasts (see ref. 33). Therefore, the broadband s-SNOM image alone cannot directly reveal the actual image of the field distributions of the discs and the frequency dependence of the hBN layer. For this, monochromatic s-SNOM images would be required; however, currently such sources are not readily available within this spectral range.

To determine the actual imaging response through the hBN layer at each individual wavelength, we performed Fourier transform infrared nanospectroscopy (nano-FTIR)[11,31] along a line scan (dashed line in Fig. 3c) across two discs. At each pixel, nano-FTIR delivers a full IR spectrum recorded at the spatial resolution of the probing tip (~50 nm). Thus we obtain s-SNOM signals $s_2(\omega, x)$ as a function of the frequency $\omega$ and the spatial position $x$.[11,16,31] This hyperspectral imaging allows the extraction of detailed spatial line profiles at various frequencies as shown in Fig. 3g. For frequencies within the lower Reststrahlen but $\omega > 780$ cm$^{-1}$, two peaks are observed for each disc (see the typical case at $\omega = 783$ cm$^{-1}$). These peaks (with the width $d'$ ~ 0.3 μm, Supplementary Information) correspond to the edge-excited HPs. The distance $D$ between the two edge-launched peaks increases from about 0.4 μm to approximately 1.25 μm (for the right disc) when changing the frequency from 780 cm$^{-1}$ to 807cm$^{-1}$, with this distance being directly related to the frequency-dependent HP propagation angle. At frequencies $\omega < 780$ cm$^{-1}$, the two edge-launched peaks are not resolved by the hBN layer due to the very small directional angle of HPs, leading to one single broad peak observed for each disc. Based on the smallest peak-to-peak separation in our results being approximately 0.4 μm at $\omega = 780$ cm$^{-1}$ ($\lambda =$ 12.8 μm), this corresponds to a sub-diffractional resolution of about $\lambda/32$ (also see Supplementary Information).

For the comparison of the theoretically predicted propagation angles (Fig. 1a) and those experimentally derived from the s-SNOM measurements, we plot the experimentally determined $D$ (blue curve and data) and corresponding directional angles $\theta$ (red curve and data) of the HPs as functions of $\omega$ in Fig. 3h. The theoretical result [from $D(\omega) = d + 2h \tan\theta(\omega)$] is also shown for comparison. Good quantitative agreement is found between the experiments and theory. This verifies the predicted tuning range of the propagation angle of the HPs to be about 35º–70º and the tunable ratio ($D/d$) to range from 1.39 to 4.2. However, certain discrepancies are still found in this comparison, and we do not observe multiple peaks originating from the multi-reflection of the HPs within the hBN slab.

**Experimental results in Type II hyperbolic band**

In order to extract the near-field frequency-dependent imaging and HP propagation angles, near-field imaging experiments were also performed in the Type II hyperbolic band by using a line-tunable quantum cascade laser (QCL, spanning from 1310 to 1430 cm$^{-1}$). This monochromatic laser allows us to present the HP-reconstructed imaging phenomenon in a more intuitive way compared to the case with the broadband laser. First, we performed the s-SNOM measurements to image a Au strip (about 1 μm long, 100 nm wide, as sketched in Fig. 4a) that is covered by the 0.15-μm hBN layer (AFM topography in Fig. 4b). In comparison to the circular discs, this rectangular object can avoid the potential confusion caused by the HP cones (circular cross section) for understanding the imaging results. Because the deeply sub-diffractional width (< λ/70) of the stripe, no optical information cannot be recorded at the frequency outside the hyperbolic band, as shown in Fig. 4c. In contrast, we clearly observe enlarged optical patterns formed by the HP (Figs. 4d-f). Qualitatively, these elliptical, rectangular patterns carry and reflect the shape information of the original object. Quantitatively, from the measured width and length of the patterns, in conjunction with the known HP propagation angle, we can extract the geometric size of the investigated object. Using this approach, we estimate the width of the stripe is about 0.12 μm, and the estimated length is about 1.1 μm, which is consistent with SEM measurements of the stripe prior to hBN exfoliation. We note that for the cases at the frequencies that HPs have smaller diffractional angle (namely, the HP cone with the very small diameter), the reconstructed image will be similar to the one-to-one near-field superlensing[29,30,31]. Therefore, in such HP-based imaging phenomenon, it has two frequency-dependent operation modes: the enlarged reconstruction and the one-to-one superlensing.

As previously discussed, we also experimentally demonstrated that the diameter of the HP cone increases with decreasing frequency, which is inverted with respect to the Type I band. These two distinct frequency dependences, found in two spectral regimes through imaging of the sub-diffractional objects, verify again that the two types of hyperbolic dispersion easily exist in the hBN crystal, without the need for fabricating different HMM structures.

In addition to the single structure, we also investigated periodic arrays of nanostructures. We imaged the arrays of the Au nanodiscs (5 × 5 array, sketched in Fig. 5a) with different separations. The diameter of the discs was fixed at 0.3 μm. The gap separation between two discs varied from g = 0.1 μm nm to g = 1 μm. The experimental results are shown in Figs 5d-f. We observed the complicated overlapping of the HP cone launched by each disc. However, with these sub-diffractional HPs, we have the opportunity to distinguish the deeply sub-diffractional structures that are not readily seen in either AFM topography or near-field imaging at frequencies outside of the hyperbolic regime (see the comparison of Figs. 5b-d).

**HPs for sub-diffractional focusing and selective waveguiding**

In addition to the already proven high-resolution imaging with tunable enlargement of the outline of sub-diffractional objects using only a simple slab of hBN, the highly directional nature of HPs can also result in sub-diffractional focusing behavior. When imaging the nanodiscs with a large diameter of 0.75 μm, we observe the focusing spot with the width of about 0.175 μm (~ λ/40) at $\omega$ = 1430 cm$^{-1}$ (see Fig. s4 in Supplementary information). This is because all the HP cones launched by the nanodisc superimpose into the center point, leading to the concentration of the light. We also investigate this focusing effect with the broadband laser in the Type I hyperbolic band. However, the obtained results are also the superposition of all the frequency components, which are not intuitive. We also note that this super-focusing effect is independently shown and discussed in Ref. 34, which was performed concurrently with the work discussed here.

Both the enlarged imaging and the super-focusing are based on the volume-confined, frequency-dependent propagation angle of HPs in hBN. We can envision other potential disruptive technologies based on the great potential of the highly directional HPs. First, this frequency-selective waveguiding could be useful for photonic switching or computing, infrared filtering, or various other nanophotonic applications. Another potential application is realized in the form of an ultra-compact subwavelength spectrometer. A natural hBN layer should allow for the spatial separation or filtering of incoming broadband light into different wavelength channels, much like a grating, which could then be detected by subwavelength IR detector pixels. This particular spectrometer configuration could also be used for chemical and biological detection schemes, in the form of spatially-resolved infrared spectroscopy. Under broadband mid-infrared illumination the HPs could carry the vibration (or absorption) information of molecules in contact with the surface, dispersing the spectral information at different angles, enabling them to be spatially resolved by a near-field intensity detector (like the s-SNOM tip) without the need of spectrometers.

**Summary and outlook**

Our work along with that of ref. 34, demonstrate the complete hyperbolic imaging response of hBN and its potential for improving the near-field imaging of deeply embedded objects[35,36] in both the Type I lower and Type II upper Reststrahlen bands, respectively. More specifically, we reveal the hyperbolic nature of the hBN layer for near-field waveguiding, imaging, focusing and its dependence upon the operational frequency. Although all our results are restricted in the near field, we also expect that these novel findings will benefit far-field imaging by introducing specific geometric designs, such as circular or wedge-shaped hyperlenses[2,3]. Furthermore, as a van der Waal's crystal[37], hBN lends itself to incorporation on non-planar and flexible substrates more amenable to true hyperlensing methodologies[4]. The realization of a naturally occurring, Type I and II hyperbolic media enable new opportunities for nanophotonics that go beyond sub-diffractional near-field imaging and potential hyperlensing. Due to the similar material anisotropy present in other polar dielectric van der Waals crystals[37], such as $MoS_2$ or $WS_2$, the natural hyperbolic response of hBN may be general to the entire class of polar two-dimensional crystals, thus expanding the potential spectral range of this behavior from the mid-IR into the single digit THz spectral region[17,24].

**Methods Summary**

Sample preparation. The gold nanostructures used for the near-field experiments were fabricated on a 1-μm-thick intrinsic silicon substrate using electron beam lithography into a bilayer PMMA resist. The nanostructures varied in size from 0.2-1 μm in diameter and in arrays with 0.1-1 μm edge-to-edge gaps. A standard liftoff procedure was used following the thermal evaporation of Cr (5 nm)/Au (30 nm) metallization.

Hexagonal BN crystals were grown using the high-pressure/high-temperature method[38,39]. The standard exfoliation process was used to randomly deposit hBN flakes of various thicknesses onto a PMMA/PMGI bilayer spun on a separate silicon substrate. Here the PMMA layer played the role of the flake carrier membrane and the PMGI served as a sacrificial lift-off layer later dissolved by TMAH solution (MICROPOSIT® MF®-319). Atomic force microscopy was utilized to select specific flakes with both sufficient thickness and lateral size for the imaging experiments. The PMMA carrier membrane with an appropriate hBN flake was lifted-off from the substrate and put onto the supportive metal ring held by a home-made micromanipulator. With the help of the micromanipulator, the hBN flake was aligned and transferred face down onto the predefined gold nanostructure by releasing the carrier membrane from the metal ring. Following the transfer, the sample with the carrier membrane on was heated to 130°C for about 10 min to soften the PMMA membrane and improve the adhesion of hBN to the underlying nanostructures and silicon substrate. After that the carrier membrane was dissolved in acetone leaving the hBN flake covering the entire array of nanostructures. To improve the adhesion, an ultrasonic clean in acetone and isopropyl alcohol with subsequent oxygen plasma clean was performed on the silicon substrate prior to the hBN transfer. More details of this transfer technique are given in Ref. 40.

Infrared s-SNOM measurements. An s-SNOM (commercially available, Neaspec GmbH) system was used to simultaneously measure the optical near fields and topography. The laser system used in Type I hyperbolic band was developed by the Fraunhofer ILT[32,33]. It consists of a commercially available ps-laser as the pump source and two subsequent nonlinear converter steps to cover the mid-IR range. The peak wavelength is continuously tunable from $\omega = 625$ cm$^{-1}$ (~16 µm) to $\omega = 1100$ cm$^{-1}$ (~ 9 µm) with bandwidths of some tens to more than hundred wavenumbers. At a repetition rate of 20 MHz and pulse duration of 10 ps, the system provides an average power of up to 10 mW. To address the lower Type-I hyperbolic region of the hBN, the peak position of the laser spectrum in our measurements was set to be at around $\omega = 790$ cm$^{-1}$ (12.7 µm) with a FWHM of about 90 cm$^{-1}$. To suppress the far-field background contribution and solely measure the near-field contribution, the optical signal s was demodulated at higher harmonics of the oscillation frequency $\Omega$ of the cantilever (in our case $2\Omega$ for the broadband s-SNOM). Nano-FTIR spectra were obtained by constantly moving the mirror in the reference arm of the Michelson interferometer, recording the resulting interferograms and their corresponding complex Fourier transformation[11,33]. In contrast to conventional far-field FTIR, this setup allows to record spectral information with a spatial resolution of down to several tens nm. For 2D imaging, the position of the reference mirror was fixed to be at the position about $\lambda/8$ away from the maximum of the interference signals. This allows for a visualization of even small spectral changes[31]. The extracted line profiles shown in Fig. 3g were numerically smoothed by using a fast Fourier Transform (FFT) smoothing with 3 adjacent pixels. This smoothing does not improve the resolution, but rather leads to a conservative estimation of resolution (details in Supplementary Information). The monochromatic QCL used in the Type II band (Fig. 4 and 5) is commercially available from the Daylight Solution (the demodulation $3\Omega$ for the monochromatic s-SNOM, $s_3$).

Numerical simulations. 2D simulations (Fig. 1) were carried out by the finite-element software COMSOL Multiphysics. A plane-wave illumination was set by using scattering boundary condition. The surrounding boundaries used perfectly matched layer absorbing boundary conditions. 3D simulations (Fig. 2) were done by using CST Microwave Studio™. Open boundary conditions were used. We also checked different mesh sizes to make sure that all the simulations reach proper convergence. The dielectric data of hBN used in all the simulations are extracted from far-field FTIR measurements[10].

**Acknowledgments**

This work was supported by the Excellence Initiative of the German federal and state governments, the Ministry of Innovation of North Rhine-Westphalia, the DFG under SFB 917 and the Korean Defense Acquisition Program Administration and the Agency for Defense

Development as a part of a basic research program under the contract UD110099GD. Funding for JDC was provided by the NRL Nanoscience Institute and was carried out at the University of Manchester through the NRL Long-Term Training (Sabbatical) Program. A.K. and K.S.N. acknowledge support from the Engineering and Physical Sciences Research Council (UK), The Royal Society (UK), European Research Council and EC-FET European Graphene Flagship.

**References**

1. Poddubny, A., Iorsh, I., Belov, P. & Kivshar, Y. Hyperbolic metamaterials. *Nature Photon.* **7**, 948–957 (2013).

2. Jacob, Z., Alekseyev, L. V. & Narimanov, E. Optical hyperlens: far-field imaging beyond the diffraction limit. *Opt. Express* **14**, 8247–8256 (2006).

3. Salandrino, A. & Engheta, N. Far-field subdiffraction optical microscopy using metamaterial crystals: theory and simulations. *Phys. Rev. B* **74**, 075103 (2006).

4. Liu, Z., Lee, H., Xiong, Y., Sun, C. & Zhang, X. Far-field optical hyperlens magnifying sub-diffraction-limited objects. *Science* **315**, 1686 (2007).

5. Hoffman, A. J. *et al*. Negative refraction in semiconductor metamaterials. *Nature Mater.* **6**, 946–950 (2007).

6. Yao, J. *et al*. Optical negative refraction in bulk metamaterials of nanowires. *Science* **321**, 930 (2008).

7. Krishnamoorthy, H. N. S., Jacob, Z., Narimanov, E., Kretzschmar, I. & Menon, V. M. Topological transitions in metamaterials. *Science* **336**, 205–209 (2012).

8. Ishii, S., Kildishev, A. V., Narimanov, E., Shalaev, V. M. & Drachev, V. P. Sub-wavelength interference pattern from volume plasmon polaritons in a hyperbolic medium. *Las. Photon. Rev.* **7**, 265–271 (2013).

9. Yang, X., Yao, J., Rho, J., Yin, X. & Zhang, X. Experimental realization of three-dimensional indefinite cavities at the nanoscale with anomalous scaling laws. *Nature Photon.* **6**, 450–454 (2012).

10. Caldwell, J. D. *et al.*, Sub-diffraction, volume-confined polaritons in the natural hyperbolic material: hexagonal boron nitride. *Nat. Comms.* 5, 5221 (2014).

11. Dai S. *et al.* Tunable phonon polaritons in atomically thin van der Waals crystals of boron nitride. *Science* **343**, 1125 (2014).

12. Prokes, S.M. *et al.* Hyperbolic and Plasmonic Properties of Silicon/Ag aligned Nanowire Arrays, *Opt. Express* **21**, 14962 (2013).

13. Khurgin J. B & Boltasseva A. Reflecting upon the losses in plasmonics and metamaterials. MRS Bull. **37**, 768-779 (2012).

14. West, P. R., Ishii, S., Naik, G. V., Emani, N. K., Shalaev, V. M., & Boltasseva, A. Searching for better plasmonic materials. *Las. Photon. Rev.* **4**, 795 (2010).

15. Tassin, P., Koschny, T., Kafesaki, M., & Soukoulis, C. M. A comparison of graphene, superconductors and metals as conductors for metamaterials and plasmonics. *Nature Photon.* **6**, 259-264 (2012).

16. R. Hillenbrand, T. Taubner, & F. Keilmann, Phonon-enhanced light–matter interaction at the nanometre scale. *Nature* **418**, 159 (2002).

17. Caldwell, J. D., Lindsay, L., Giannini, V., Vurgaftman, I., Reinecke, T. L., Maier, S. A., & Glembocki, O. J. Low-loss, infrared and terahertz nanophotonics using surface phonon polaritons. Nanophotonics, http://dx.doi.org/10.1515/ nanoph-2014-0003 (2014).

18. Caldwell, J. D. et al. Low-loss, extreme sub-diffraction photon confinement via silicon carbide surface phonon polariton nanopillar resonators. *Nano Lett.* **13**, 3690-3697, (2013).

19. Wang T., Li P., Hauer B., Chigrin D. N. & Taubner T. Optical properties of single infrared resonant circular microcavities for surface phonon polaritons. *Nano Lett.* **13**, 5051 (2013).

20. Chen Y., Francescato Y., Caldwell J.D., Giannini V., Mass T., Bezares F. J., Taubner T., Kasica R., Hong M., Maier S. A. Tracking the role of proximity and size for localized surface phonon polariton resonators, ACS Photonics, **1**, 718-724 (2014).

21. Xu X. G. *et al.* One-dimensional surface phonon polaritons in boron nitride nanotubes. *Nat. Comms.* **5,** 4782(2014).

22. Da Silva R. E., Macedo R., Dumelow T., Da Costa J. A. P., Honorato S. B. & A. P. Ayala. Far-infrared slab lensing and subwavelength imaging in crystal quartz. *Phy. Rev.* B **86**, 155152 (2012).

23. Fonoberov V. A. & Balandin A. A. Polar optical phonons in wurtzite spheroidal quantum dots: theory and application to ZnO and ZnO/MgZnO nanostructures. J. Phys.: Condens. Matter **17**, 1085 (2005).

24. Thompson, D. W., De Vries, M. J., Tiwald, T. E., & Woollam, J. A. Determination of optical anisotropy in calcite from ultraviolet to mid-infrared by generalized ellipsometry. *Thin Solid Films*, **313**, 341-346 (1998).

25. Sun, J., Litchinitser, N. M., Zhou, J., Indefinite by nature: from ultraviolet to terahertz. *ACS Photonics* **1**, 293 (2014).

26. Zhang Y., Fluegel B., Mascarenhas A., Total negative refraction in real crystals for ballistic electrons and light *Phys. Rev. Lett.* **91**, 157404 (2003).

27. Chen X. L., He M., Du Y. X., Wang W. Y., Zhang D. F., Negative refraction: an intrinsic property of uniaxial crystals *Phys. Rev. B* **72**, 113111 (2005).

28. Geick, R., Perry, C. H., & Rupprecht, G. Normal modes in hexagonal boron nitride. *Phy. Rev.* B, **146**, 543 (1966).

29. J. B. Pendry, Negative refraction makes a perfect Lens. *Phys. Rev.* Lett. **85**, 3966 (2000).

30. Fang N. *et al.* Sub-diffraction-limited optical imaging with a silver superlens. *Science* **308**, 534 (2005).

31. Taubner T., Korobkin D., Urzhumov Y., Shvets G. & Hillenbrand R. Near-field microscopy through a SiC superlens. *Science* **313**, 159 (2006).

32. Wueppen, J., Jungbluth, B., Taubner, T., & Loosen, P. Ultrafast tunable mid IR source. In Infrared, Millimeter and Terahertz Waves (IRMMW-THz), 36[th]-International Conference, IEEE, pp. 1-2 (2011).

33. Bensmann, S. *et al.* Near-field imaging and spectroscopy of locally strained GaN using an IR broadband laser. *Opt. Express* **22**, 22369-22381 (2014).

34. Dai S. *et al.* Subdiffractional focusing and guiding of polaritonic rays in a natural hyperbolic material. Submitted (2014).

35． Taubner T., Keilmann F., Hillenbrand R. Nanoscale-resolved subsurface imaging by scattering-type near-field optical microscopy. *Opt. Express* **13**, 8893-8899 (2005).

36． Li, P., Wang, T., Böckmann, H., & Taubner, T. Graphene-enhanced infrared near-field microscopy. *Nano Lett.* **14**, 4400-4405 (2014).

37． Geim, A. K. & Grigorieva, I. V. Van der Waals heterostructures. *Nature* **499**, 419-425, (2013).

38． Taniguchi, T. & Watanabe, K. Synthesis of high-purity boron nitride single crystals under high pressure by using Ba-BN solvent. *J. Cryst. Growth* **303**, 525-529, (2007).

39． Watanabe, K., Taniguchi, T. & Kanda, H. Direct-bandgap properties and evidence for ultraviolet lasing of hexagonal boron nitride single crystal. *Nature Materials* **3**, 404-409, (2004).

40. A. V. Kretinin *et. al*, Electronic Properties of Graphene Encapsulated with Different Two-Dimensional Atomic Crystals. *Nano Lett.* **14**, 3270-3276, (2014).

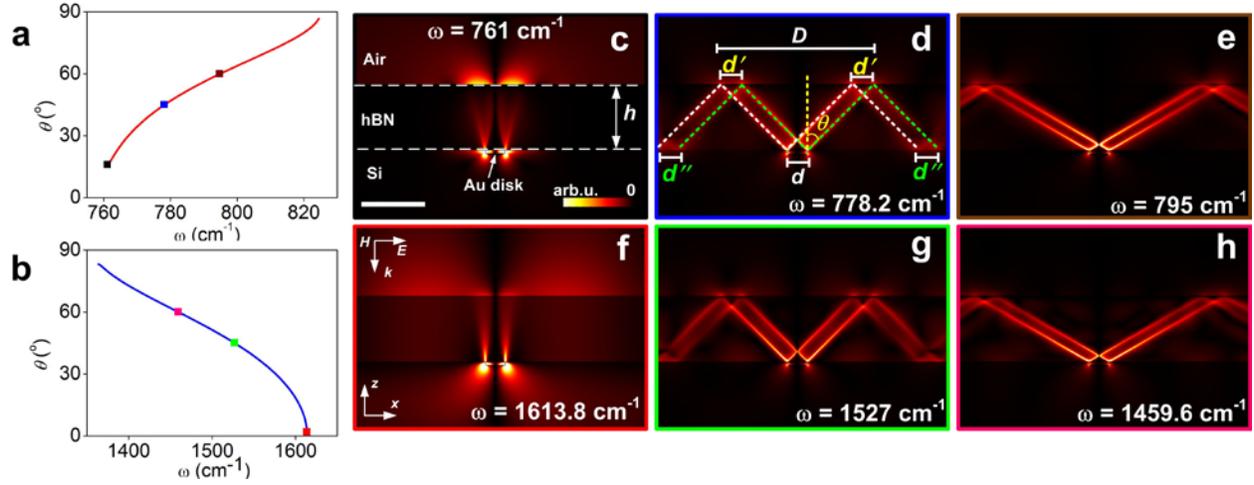

**Figure 1. Frequency-dependent directional angles of the HPs propagating inside the hBN.** Solid lines in **a** and **b,** the critical angle $\theta$ of the HPs as a function of the frequency $\omega$ in the Type-I (760 < $\omega$ < 825 cm$^{-1}$, $\varepsilon_t$ < 0 and $\varepsilon_z$ > 0) and Type-II (1360 < $\omega$ < 1610 cm$^{-1}$, $\varepsilon_t$ > 0 and $\varepsilon_z$ < 0) hyperbolic bands of the hBN. **c-h,** Simulated electric-field distribution ($|E_z|$) at various frequencies. The directional angles evaluated from these simulations are plotted in **a** and **b** (color dots) for comparison: $\theta$ = 16° (**c**, at $\omega$ = 761 cm$^{-1}$), $\theta$ = 2° (**f**, at $\omega$ = 1613.8 cm$^{-1}$), $\theta$ = 45° (**d** and **g**, at $\omega$ = 778.2 cm$^{-1}$ and $\omega$ = 1527 cm$^{-1}$) and $\theta$ = 60° (**e** and **h**, at $\omega$ = 795 cm$^{-1}$ and $\omega$ = 1459.6 cm$^{-1}$). The scale bar in **c** indicates 1 μm.

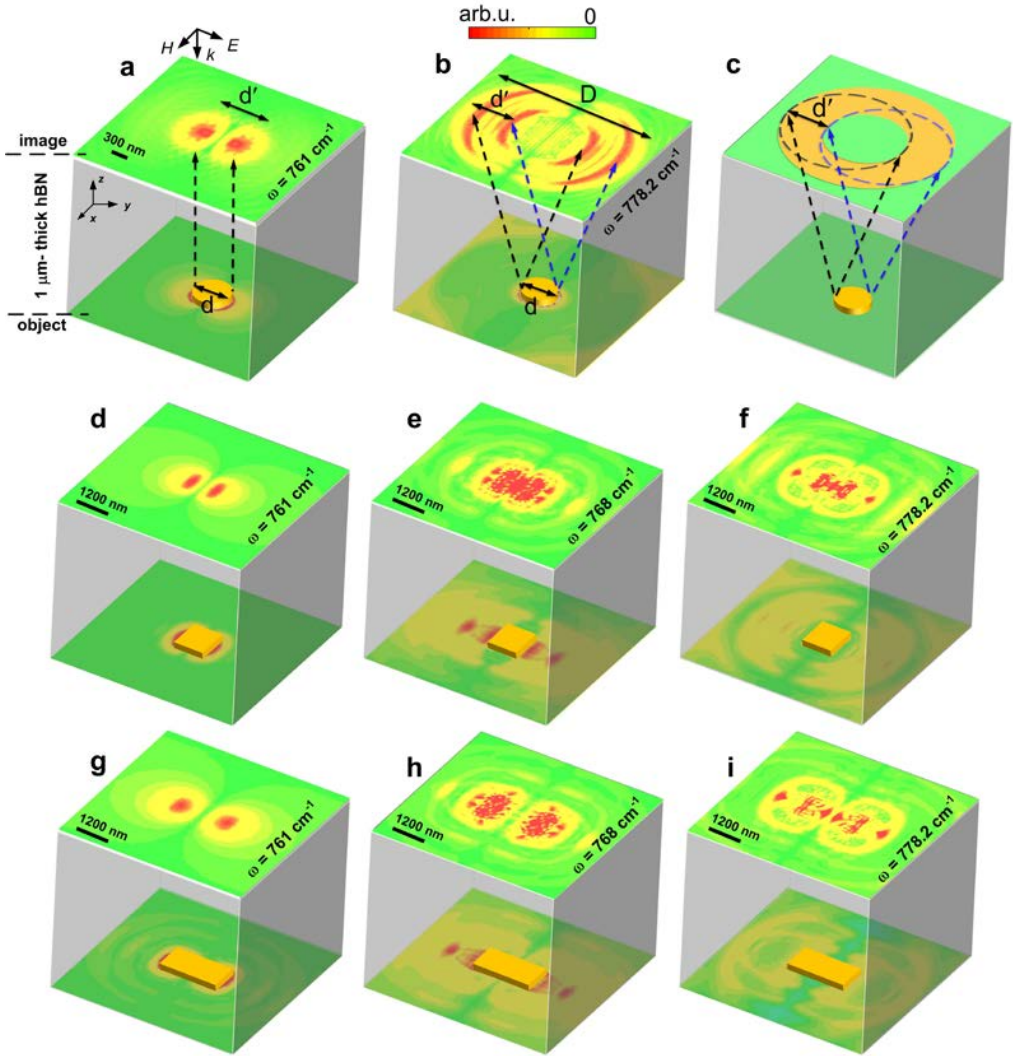

**Figure 2. Three dimensional simulations of imaging different structures through the hBN layer. a** and **b**, 3D simulations of imaging a gold disc (0.6-μm diameter) below the 1-μm-thick hBN layer (scale bar 300 nm). Simulated electric-field distributions ($|E_z|$) taken at top and bottom surface of the hBN layer for imaging, **a,** at $\omega = 761$ cm$^{-1}$. **b,** at $\omega = 778.2$ cm$^{-1}$ show the frequency-dependent transition between perfect imaging and enlarged imaging. **c,** the sketch of the mechanism of the enlargement observed in **b**. (In this sketch we do not consider the influence of the illumination polarization). **d-e**, $|E_z|$-distributions of imaging a gold square (1-μm length, 50-nm height). **f-i**, $|E_z|$-distributions of imaging a gold bar (1-μm width, 2-μm length, 50-nm height). All these images show the frequency-dependent transition between perfect imaging and enlarged imaging of the geometric outline of the structures. The z-axis in all the images is not to scale for better visualization.

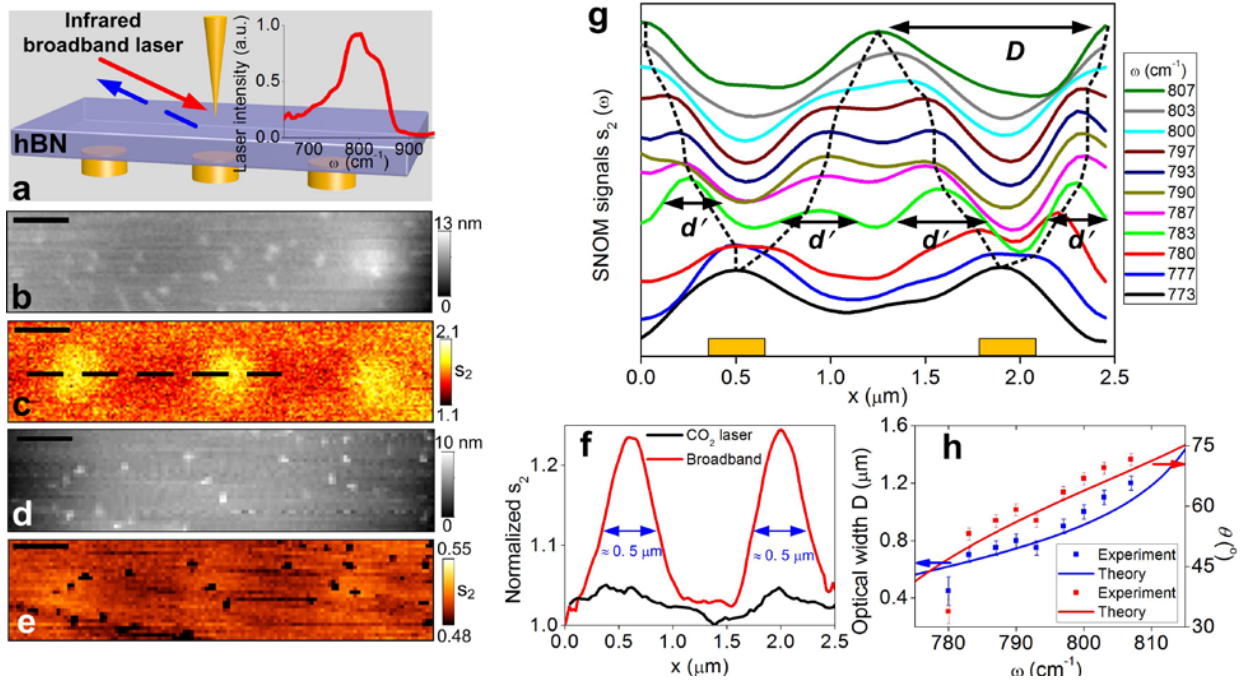

**Figure 3. Experimental demonstration of super-resolution imaging with tunable HPs in the Type I band. a,** Sketch of the experimental set-up. The right inset is the normalized laser spectrum of the used mid-infrared broadband laser. **b,** The AFM topography taken at the top surface of the 0.15-μm-thick hBN hyperlens. **c,** The 2D infrared optical images taken with the broadband laser. **e,** The control infrared image taken with a $CO_2$ laser at $\omega = 952$ cm$^{-1}$ that is out of the hyperbolic region of the hBN. The small black dots in the image are caused from topographic features (corresponding topographic image shown in **d**). Scale bars: 0.5 μm. **f,** Detailed profiles of the s-SNOM signals across two neighboring discs (along the line marked in **e**, averaged over 5 scan lines) for the cases using the broadband laser (red line) and the $CO_2$ laser (black line), respectively. Both profiles are normalized to their respective minimum values outside the discs. The broadband imaging shows much stronger contrasts for the discs. **g,** Detailed Nano-FTIR line profiles at various frequencies. Dashed line marks the position variations of the peak of edge-launched HPs. **h,** Optical widths and corresponding directional

angles of the HPs evaluated from the experimental results (dots) in comparison with the calculated results (solid lines). The error bars result from the spatial pixel (~50 nm) in nano-FTIR measurements.

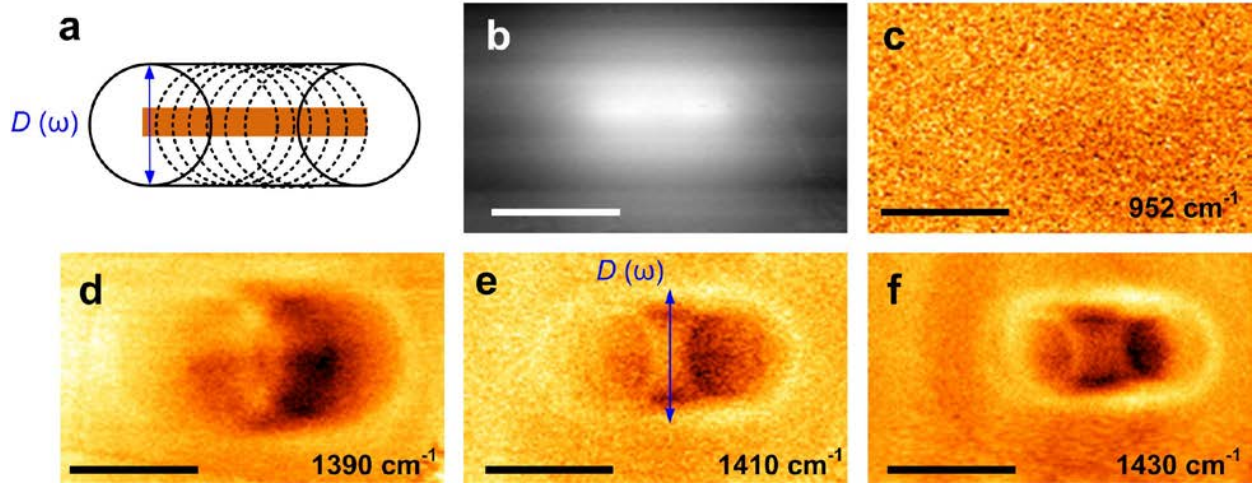

**Figure 4. Experimental demonstration of enlarged imaging of a Au stripe with tunable HPs in the Type II band. a,** Sketch of the imaged object - a Au stripe (not to scale), located below the 0.15 μm thick hBN flake. The length $l$ of the stripe is about 1 μm, and the width $w$ is around 0.1 μm. The dashed rings result from launched HPs from the edges of the stripe. These HPs form an enlarged outline that reveals the object information. **b,** AFM topography taken from the hBN top surface. **c,** the s-SNOM image taken at $\omega = 952$ cm$^{-1}$ outside the hyperbolic band. No optical feature is observed from the underlying structure. **d - f,** Spectroscopic imaging the stripe in the Type II hyperbolic band. The bright outlines found in the images are formed by the HP cones, which are enlarged compared to the original structure. From the measured $D(\omega)$, we can estimate the width $w$ of the stripe (using the relationship $D(\omega)= w + 2h \tan\theta(\omega)$) is about 0.12 μm, and the estimated length is about 1.1 μm. These results verify that from imaging the enlarged outline, we are able to reveal the object information due to the frequency-dependent directivity of HPs. The scale bars indicate 1 μm.

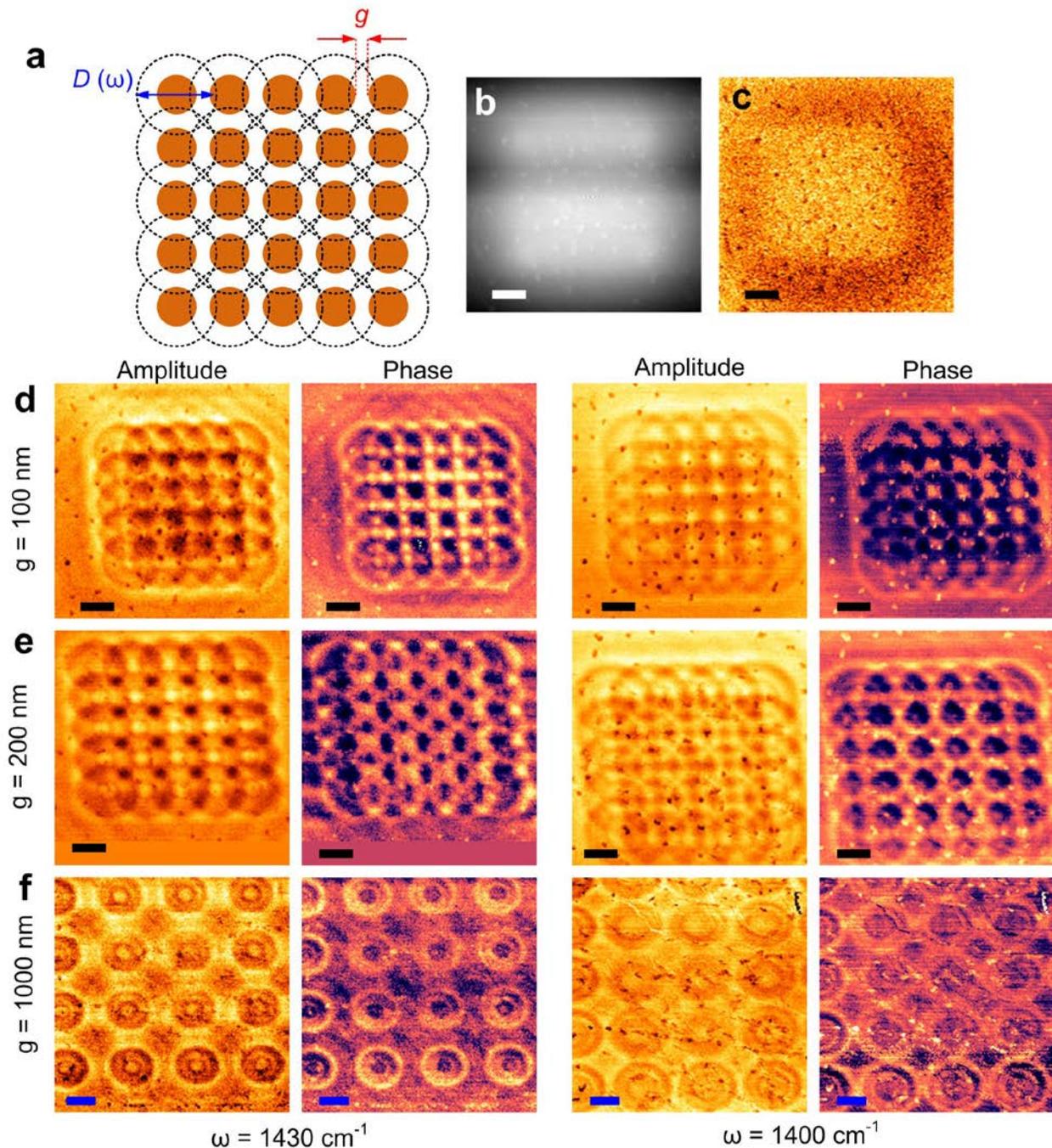

**Figure 5. Experiments of imaging the arrays of Au nanodiscs with HPs in the Type II band.**

**a,** Sketch of the arrays of the Au nanodiscs. The diameter of the discs is fixed to be 300 nm. The gap separation between two discs varies from g = 0.1 μm to g = 1 μm. The dashed rings indicate the launched HPs. **b,** the AFM topography of the array with g = 0.1 μm, taken from the hBN top surface. **c,** the s-SNOM image of the array with g = 100 nm taken at $\omega = 952$ cm$^{-1}$ outside the hyperbolic band. No optical feature is probed for resolving the structure. **d - f,** Near-field images (amplitude and phase) of the arrays at two different frequencies of 1430 cm$^{-1}$ and 1400 cm$^{-1}$ inside the Type II hyperbolic band. Obviously, the launched HPs help to reveal the arrays. In **e**), color spaces are added in the amplitude and phase images taken at 1430 cm$^{-1}$, for keeping the square shape. All the scale bars indicate 0.5 μm.

# Supplementary information for "Hyperbolic phonon-polaritons in boron nitride for near-field optical imaging and focusing"

Peining Li[1], Martin Lewin[1,2], Andrey V. Kretinin,[3] Joshua Caldwell[4]*, Kostya S. Novoselov,[3] Takashi Taniguchi,[5] Kenji Watanabe,[5] Fabian Gaussmann[2], Thomas Taubner[1,2]*

1 Institute of Physics (IA), RWTH Aachen University, Aachen 52056, Germany

2 Fraunhofer Institute for Laser Technology ILT, 52074 Aachen, Germany

3 School of Physics and Astronomy, University of Manchester, Oxford Rd, Manchester, UK

4 U.S. Naval Research Laboratory, 4555 Overlook Ave, S.W., Washington, D.C. USA

5 National Institute for Materials Science, 1-1 Namiki, Tsukuba, Ibaraki 305-0044, Japan

*Correspondence to: joshua.caldwell@nrl.navy.mil and taubner@physik.rwth-aachen.de

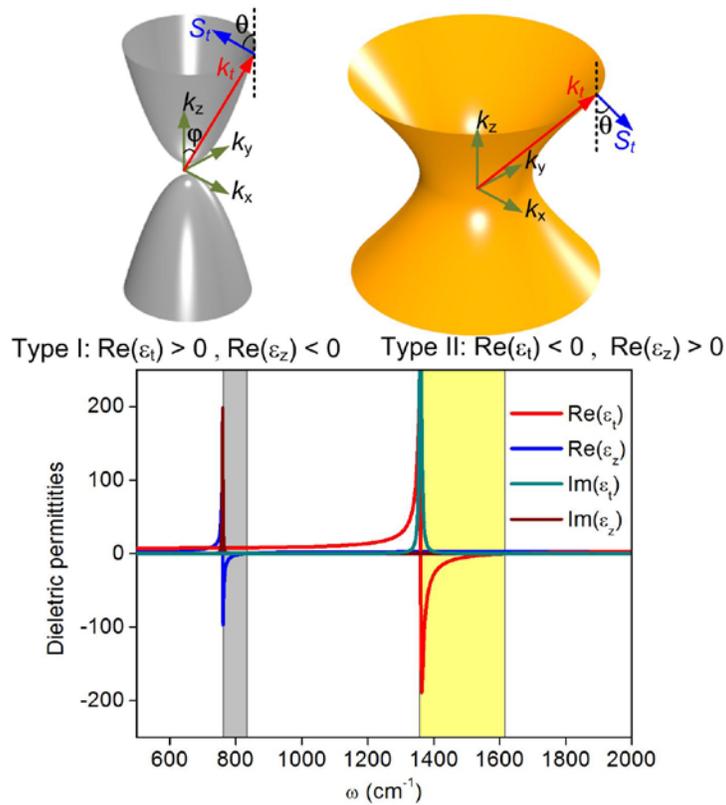

**Figure S1. Hyperbolic dispersion in a hBN layer.** Top, two types of hyperbolic dispersions (the Type I for Re($\varepsilon_t$) < 0 and Re($\varepsilon_z$) > 0 and the Type II for Re($\varepsilon_t$) > 0 and Re($\varepsilon_z$) < 0) and their corresponding vector diagrams. $\varphi$ is the angle between the wavevector $k_t$ and $z$ axis. $\theta$ is the angle between the Poynting vector $S_t$ and $z$ axis. For large wavevectors ($k_x \gg k_0$, $k_0 = 2\pi/\lambda$), $\varphi + \theta \approx \pi/2$. Bottom, in-plane and out-plane dielectric permittivities ($\varepsilon_t$, $\varepsilon_z$) of hBN, taken from the ref.10. Two hyperbolic regions are marked in gray and yellow, respectively.

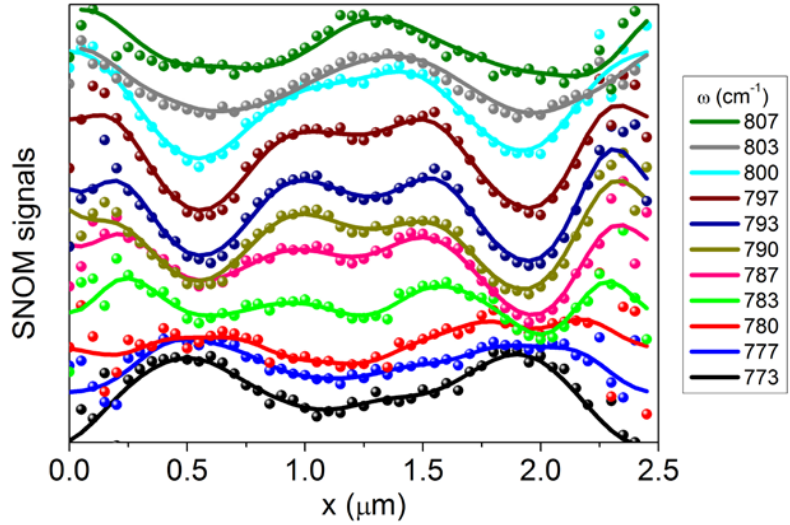

Figure S2. Experimental data of nano-FTIR line profiles with (solid lines) and without (color dots) the numerical smoothing.

**Determination of the resolution.** As mentioned in the main text, the smallest peak width of launched HPs that we can resolve in the Nano-FTIR line scans is about 0.3 μm at $\omega = 783$ cm$^{-1}$ (about 12. 8 μm, see Fig. S2), corresponding to a deep subwavelength scale of $\lambda/42$. However, for defining the optical spatial resolution, we consider the smallest resolvable peak-to-peak separation of about 0.4 μm (Namely, $\lambda/32$ resolution).

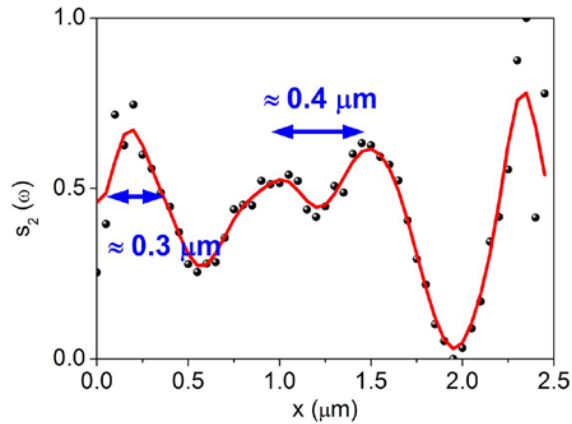

Figure S3. Experimental data of nano-FTIR line profiles with (solid lines) and without (black dots) the numerical smoothing at $\omega = 783$ cm$^{-1}$ (about 12. 8 μm).

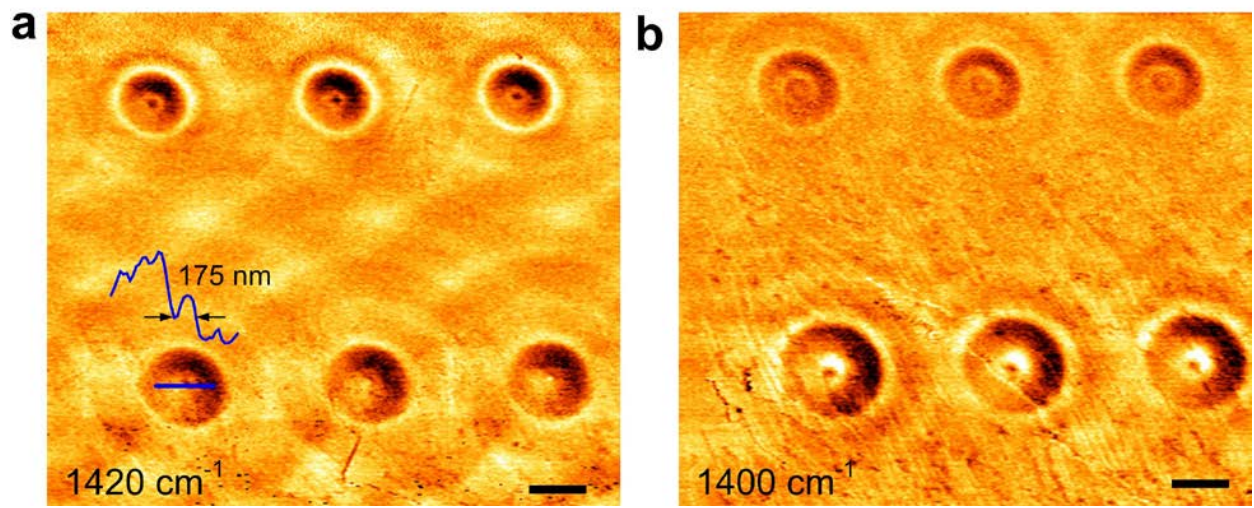

Figure s4. Experimental demonstration of near-field optical focusing and waveguiding of HPs. Near-field images of the nanodiscs (top three with diameter 500 nm, bottom three with diatmeter 750 nm) **a,** at 1420 cm$^{-1}$ **b,** at 1400 cm$^{-1}$. Clear focusing and interference of HPs are found. The scale bars indicate 1000 nm. From the line profile across the focusing spot (shown in **a**, along the blue line), a width of about 175 nm (~ $\lambda/40$) of the spot are obtained.

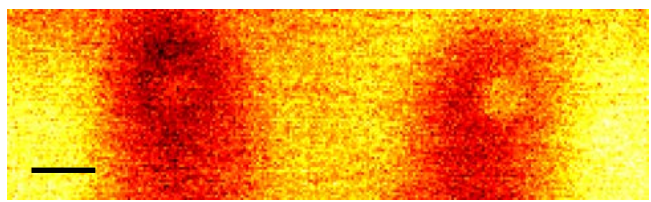

Figure S5. Experimental results of imaging the nanodiscs with the diameter of 500 nm with the broadband laser in the Type I hyperbolic band. The scale bar indicates 1000 nm.

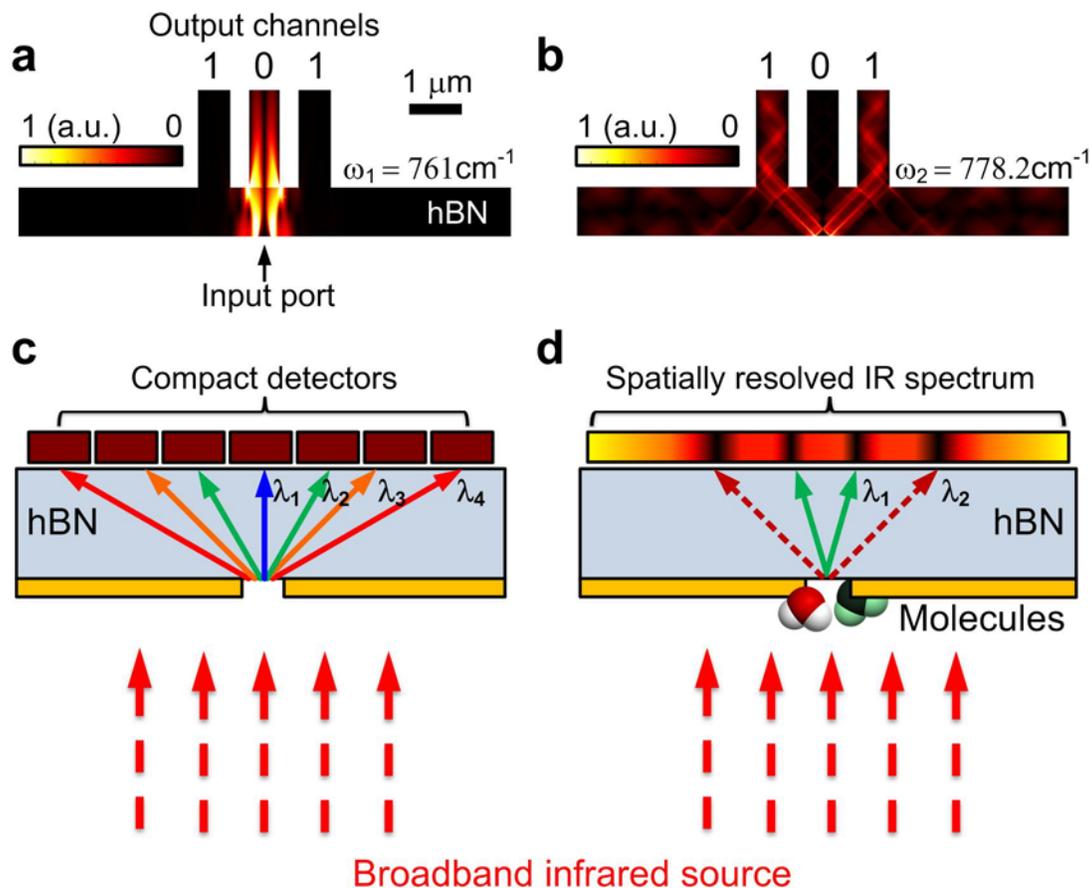

Figure s6. Further application proposals using the tunable highly directional HPs. Simulated $|E_z|$-distributions inside an hBN waveguiding structure (**a**) at $\omega = 761$ cm$^{-1}$, (**b**) at $\omega = 778.2$ cm$^{-1}$. **e**, Sketches of an ultra-compact hBN-based infrared spectrometer. The thin hBN layer allows spatially splitting different wavelength components of the incoming infrared illumination. **f**, Sketches of hBN-based spatially-resolved infrared spectroscopy for different molecules. The dark colors in the spectrum represent the absorption by the molecules (sketch).

Figure s6a shows numerical simulations of HPs propagating inside an hBN waveguide structure (a horizontal hBN-layer connected with three vertical hBN-channels) at a frequency of $\omega = 762$ cm$^{-1}$. Due to the small propagation angle, the electromagnetic fields primarily propagate into the inner '0'-channel. When tuning the frequency to $\omega = 778.2$ cm$^{-1}$ (Fig. s6b), the fields propagate with an angle of ~45º and therefore transmit into the two outer '1'-channels. More complicated geometries could also expand such approaches beyond a binary routing into more sophisticated multi-channel systems. Consequently, this frequency-selective waveguiding could be useful for photonic switching or computing, infrared filtering, or various other nanophotonic applications. Another potential application is realized in the form of an ultra-compact subwavelength spectrometer. As sketched in Fig. s6c, a natural hBN layer should allow for the spatial separation or filtering of incoming broadband light into different wavelength channels, which could then be detected by subwavelength IR detector pixels. This particular spectrometer configuration could also be used for chemical and biological samples, in the form of spatially-resolved infrared spectroscopy. Under broadband mid-infrared illumination the HPs could carry the vibration (or absorption) information of molecules in contact with the surface (Fig. s6d), dispersing the spectral information at different angles, enabling them to be spatially resolved by a near-field intensity detector (like the s-SNOM tip) without the need of spectrometers.